\documentclass[useAMS,usenatbib]{mnras}
\usepackage{color}
\usepackage{graphicx}
\usepackage{dcolumn}
\usepackage{bm}
\usepackage{amssymb}
\usepackage{latexsym}
\usepackage[T1]{fontenc}
\usepackage{aecompl}
\usepackage{amsmath}%
\usepackage{enumitem}
\usepackage{epstopdf}
\usepackage{booktabs}
\usepackage{setspace}
\newcommand{\be}{\begin{equation}}
\newcommand{\ee}{\end{equation}}
\newcommand{\bq}{\begin{eqnarray}}
\newcommand{\eq}{\end{eqnarray}}

\def\({\left(}
\def\){\right)}

\begin{document}

\title[GW emission and cosmic heat death]{A preliminary study about gravitational wave radiation and cosmic heat death}
%%\short{GW and cosmic heat death}

\author[Zhang, Qian, Guo, Wang, \& Li (2020)]
{Jianming Zhang $^1$, Qiyue Qian $^{1\dagger}$, Yiqing Guo $^1$, Xin Wang $^{1\ddagger}$,Xiao-Dong Li $^{1\ast}$ \\
$^1$ School of Physics and Astronomy, Sun Yat-Sen University, Guangzhou 510297, P. R. China \\
%$^2$ Departamento de F{\'i}sica, Universidad de los Andes, Cra. 1 No. 18A-10 Edificio Ip, CP 111711, Bogot{\'a}, Colombia \\
%$^3$ Department of Physics and Astronomy, Sejong University, Seoul, 143-747, Korea \\
$^\dagger$Corresponding author: qianqy5@mail2.sysu.edu.cn \\
$^\ddagger$Corresponding author: wangxin35@mail.sysu.edu.cn \\
$^\ast$Corresponding author: lixiaod25@mail.sysu.edu.cn}

\pagerange{\pageref{firstpage}--\pageref{lastpage}} \pubyear{2020}
\maketitle
\label{firstpage}

%\date{\today

\begin{abstract}
We study the role of gravitational waves (GW) in the heat death of the universe. %in the heat death.
Due to the GW emission, in a very long period, dynamical systems in the universe suffer from persistent mechanical energy dissipation, evolving to a state of universal rest and death.
With N-body simulations, we adopt a simple yet representative scheme to calculate the energy loss due to the GW emission.
For current dark matter systems with mass $\sim10^{12}-10^{15} M_\odot$,
we estimate their GW emission timescale as $\sim10^{19}-10^{25}$ years.
This timescale is significantly larger than any baryon processes in the universe, 
but still $\sim10^{80}$ times shorter than that of the Hawking radiation.
We stress that our analysis could be invalid due to many unknowns
such as the dynamical chaos, the quadrupole momentum of halos,  
the angular momentum loss, the dynamic friction,  
the central black hole accretion, the dark matter decays or annihilations, 
the property of dark energy and the future evolution of the universe.
\end{abstract}

%\keywords{ large-scale structure of universe --- dark energy --- cosmological parameters }

\begin{keywords}
gravitational waves, cosmology, astrophysics
 \end{keywords}

%{\bf Comments:

%* Use larger fonts in all figures, especially in the legend.

%* remove frames of the legends: frameon=False

%*

%* Define $\hat P_{\Delta k}(k,\mu)$, donot use the many $\int P$

%}

\section{Introduction}\label{intro}

%which has

The heat death fate of the universe, also known as the "Big Chil" or "Big Freeze",
is a conjecture suggesting that the universe would
end up with a state of no thermodynamic free energy.
In such situation the universe would be unable to sustain any process that increases entropy.

The idea of heat death was first studied by Kelvin, %William Thomson (Lord Kelvin),
who %took the theory of heat
%as mechanical energy loss in nature
%(as embodied in the first two laws of thermodynamics)
extrapolated the second law of thermodynamics of mechanical energy dissipation
%as the dissipation of mechanical energy
%and extrapolated it to the universe scale.
to a cosmic scale in the 1850s,
and found it inevitably led to "a state of universal rest and death".
%His conjecture is echoed by ,
Along with this conjecture, Helmholtz \citep{Kelvin:1852}
envisioned a stable, thermodynamic equilibrium state in the end,
stating that "the universe from that time forward
would be condemned to a state of eternal rest." 
Similar viewpoint was held by Clausius,
who believed that the universe would evolve to
"a state of unchanging death"
once its entropy reaches the maximum.
%{\bf Jiaming: Make sure the words in ``'' are%
%the original words of these great guys?}

%Helmholtz made the first explain of this theory (Helmholz,1854),
%envisioned stability in the end Comprehensive
%thermodynamic equilibrium state:
%"the universe from that time forward
%would be condemned to a state of eternal rest"
%and echoed by Clausius (Clausius. R, 1865; Clausius. R, 1868):
%the universe would be in a state of unchanging death,
%if the entropy is maximum.
%This means the establishment of heat death hypothesis.

% The discussions on the correctness of the heat death hypothesis never stops since it proposed.
% {\bf Xiaodong: left this headache paragraph to me if you do not know how to improve it...}
% Since the formulation of heat death hypothesis, the discussion on the correctness of it never stops.
% Some voices of opposition, such as ``rise and fall theory'' by L.E.Boltzman and ``Maxwell's demon'' (Bennett, 1982),
% while they didn't prove it mistaken at all.
% Conversely, the heat death ideas were widely discussed.
% Jeans (Jeans, 1930) and Eddington (Eddington, 1928; Eddington, 1931) were bringing the notion of
% the ``heat death'' into relativistic cosmology.
% Barrow and Tipler (Barrow and Tipler, 1978)
% demonstrated that the Hawking black hole evaporation process causes a vortical instability
% to develop in spatially homogeneous spacetimes, furthermore they provided a novel picture of the universal heat death.

%{\bf Xiaodong: left this headache paragraph to me if you do not know how to improve it...}
Discussions on the heat death hypothesis never stops since it was proposed.
%The heat death hypothesis was widely extended.
\cite{Jeans:1930,Eddington1931} brought the notion of
the heat death into relativistic cosmology.
\cite{Barrow:1978} demonstrated that the Hawking black hole evaporation process would cause a vortical instability
to develop in spatially homogeneous spacetimes, and furthermore provided a novel picture of the universal heat death.
Also, there were opposing viewpoints to the idea of heat death,
such as "rise and fall theory" by L.E.Boltzman and the conjecture of  "Maxwell's demon" \citep{Bennett:1982}, etc. 

With the development of modern cosmology, it becomes commonly accepted that
the ultimate fate of the universe crucially depends on its energy components.
%is tproperties of dark matter and dark energy. % is particularly critical.
Heat death is likely to happen in a forever expanding universe,
which is expected to occur if the topology of the universe is open or flat (in a matter-dominated universe), or if the dark energy component keeps dominating.
The latter possibility is currently supported by multiple cosmological observations (see \cite{ade2016planck} and the references therein).
%which reveals that two thirds of components in our universe is composed of a mysterious dark energy
%about matter(mainly by dm)
%implies either the existence of a “dark energy” component in the universe or the breakdown of
%Einstein’s gravity theory on cosmological scales

\cite{Dyson:1979} provided a comprehensive summary of physical processes related to the heat death, which included the stellar evolution, the detachment of planets from stars and stars from galaxies,  the decay of object orbits by gravitational radiation, 
the evaporation of black holes by the Hawking process, 
the liquification of all matter at zero temperature, 
the decay of matter to iron,  as well as the collapse of iron stars to neutron stars and the collapse of ordinary matter to black holes.
%These issues are not discussed in our work, and it is a worthy future investigation. 
%While we don't think so much about it, we just think about the black holes caused by gravitational waves emission and the radiation part of the black holes.
While most of these processes are of great interest and worth more detailed investigation,
in this paper, we focus on the detailed investigation of the {\it gravitational waves}.
As is mentioned in \cite{Barrow:1978}, 
``as time passes, clusters, galaxies and other stellar systems will become increasingly bound by gravitational forces 
since they will radiate away their binding energy in the form of {\it gravitational waves}''.

Gravitational waves (GWs) generated by accelerated mass are spacetime curvature disturbances 
which propagate as waves outward from the source at the speed of light
\citep{Eistein1916,Eistein1918}.
%http://www.academie-sciences.fr/pdf/dossiers/Poincare/Poincare_pdf/Poincare_CR1905.pdf
%Einstein, A (June 1916). "Näherungsweise Integration der Feldgleichungen der Gravitation". Sitzungsberichte der Königlich Preussischen Akademie der Wissenschaften Berlin. part 1: 688–696. Bibcode:1916SPAW.......688E. Archived from the original on 2016-01-15. Retrieved 2014-11-15.
%Einstein, A (1918). "Über Gravitationswellen". Sitzungsberichte der Königlich Preussischen Akademie der Wissenschaften Berlin. part 1: 154–167. Bibcode:1918SPAW.......154E. Archived from the original on 2016-01-15. Retrieved 2014-11-15.
%In 2015, advanced LIGO (LIGO, 2015)
%made the first direct detection of the GW signal GW150914,
On September 14th, 2015, the Laser Interferometer Gravitational Wave Observatory (LIGO)
made the first direct detection of a GW event \citep{PhysRevLett.116.061102},
opening a new window for us to test the predictions of general relativity 
and probe into the structures of the universe.
%was observed with
%which became a successful confirmation of
%a prediction by Einstein's theory of general relativity (GR).
%As a new way of
GWs take place in all dynamical systems, causing persistent mechanical energy dissipation in the long-term evolution of our universe, and therefore are expected to become an essential process to the heat death.
%However, the discover of gravitational wave suggests a new way of energy transformation and transfer,
%which may have an effect of the long term evolution of the universe and need to be considered.

%
GWs carry energy away from their sources and,
in the case of orbiting objects, lead to an in-spiral or the shrinking of the orbit's radius.
Before the detection made by LIGO,
pulsar timing observations over decade had shown
a gradual decay of the pulsar binary orbital period
that matched the loss of energy and angular momentum
in GW radiation predicted by general relativity;
this offered the first indirect evidence of the existence of GWs
\citep{HT1975,TW1982}.
This process ends up with the merger of the objects,
which would form a heavier compact object (e.g. a black hole) and
produce strong GW signals detectable via experiments like LIGO.

The in-spiral caused by GW emission
is a universal process of mechanical energy loss,
regardless of the nature of the objects.
%This offers dark matter
Therefore, it will become the key mechanism for dark matter systems
to reach heat death. %can reach heat death state via this process.
Unlike the baryonic systems such as gas, stars, galaxies and galaxy clusters,
dark matter can not lose mechanical energy via thermal radiation and (non-dynamical) friction.
Therefore, the basic picture regarding our thermal fate would be that dark matter systems gradually lose mechanical energy, become more compact, and finally collapse into black holes (BHs). Then BHs evaporate, and the universe eventually enters the "state of eternal rest".

%mechanical energy tend to lose and become thermal energy in many ways,
%such as friction, thermal radiation, etc.

%The mechanical energy dissipation due to GWs has been
%When masses orbit each other, which is a common kind of accelerated motion,
%the orbit energy, including their gravitational potential energy and kinetic energy,
%would gradually lost and be carried out by GW radiation,
%and dark matter is no exception.
%Thus, matter, include baryons and dark Thus matter, will gradually gather together,
%until it collapses into a black hole,
%and then continue lose mass and energy by Hawking radiation until disappear.
%This is a possible picture of how the universe come to heat death,
%and since the gravitational wave radiation is always exist,
%this shows the upper limit of the lifetime of the universe.

%The
%The development of cosmology also inspires people to think about the energy transformation about matter
%is important to decide whether the universe would come to the end of heat death.
%As usually considered, mechanical energy tend to lose and become thermal energy in many ways,
%such as friction, thermal radiation, etc.
%But as for dark matter that predominate,
%there is no electromagnetic way to dissipate, which play an important role in this transformation.

In this work we conduct a preliminary yet representative investigation on the role of GW radiation in the cosmic death.  
%We %conduct %a preliminary study about the GW radiation's role in cosmic heat death,
%estimate the upper limit of universe's lifetime by analysing the process that matter, mainly dark matter,
%gradually loses energy by gravitational wave radiation until collapsing into Schwarzschild black hole,
%then evaporates until there is no heat exchange.
In \ref{Sec:gw_emi} we briefly introduce how GW emission affects the dynamical systems. %simplify the model to get the expressions of the time in processes mentioned above.
In \ref{Sec:method}, we estimate the GW emission timescale of dark matter systems in the current universe with the aid of simulations. %the model to dark matter simulations for estimating the timescale taken.
In \ref{Sec:caveats} we discuss the caveats of our analysis as well as some closely related issues.
We summarize and conclude in \ref{Sec:conclusion}.
%Finally, we discuss some issues which may also affect the picture of the cosmic heat death in Section 4.

\section{GW emission in dynamical systems}
\label{Sec:gw_emi}
By carrying energy away, the GW radiation causes a persistent loss of mechanic energy in a system. As a gravitational effect, this process only involves the mass and dynamical properties of the system.
In the most simplified case of a two-body system with orbital velocity $v$ and mass of objects $m_1$ and $m_2$, the rate at which energy is carried away by GWs is given as follows,
%As a preliminary estimation of the upper limit of the time to reach the heat death,
%and in consideration that dark matter which predominate over matter cannot participate in electromagnetic interactions,
%we simplify the model, only consider the GW's contribution to the lose of mechanic energy.
%And to simplify it more,
%we put the massive system hold by gravity equivalent to the two-body system model for our analysis,
%and ignore the relative motion in those two part whose GW radiation is much smaller.
%Then we conducted the following calculation for two-body system.
%According to the analysis of gravitational-wave by LIGO (LIGO Scientific and VIRGO Collaborations, 2017) ,
\begin{equation}\label{eq:dEGWdt}
{d\over dt}E_{GW}={32\over 5}{G\over c^5}u^2r^4\omega^6~,
\end{equation}
where we define $u=m_1m_2 / M$, $M=m_1+m_2$,
and $\omega$ being the angular velocity.
Here the constants $G=6.67\times10^{-11}\rm m^3 / s^2 \  kg$,
$c=2.998\times 10^8 \rm m /s$.
Notice that the mechanic orbital energy takes the form of
%and orbit energy is
\begin{equation}\label{eq:Eorb}
E_{orb}={-GMu\over 2r}~.
\end{equation}
Considering the energy loss is only caused by GW radiation, we have
\begin{equation}\label{eq:dEorbdt}
{d\over dt}E_{orb}={GMu\over 2r^2}\dot{r}={-d\over dt}E_{Gw}~.
\end{equation}
Now we are ready to estimate the timescale that the system takes to form a compact object, which we denote as
$\Delta_t \equiv t_s-t_0$, where $t_0$ is the current epoch and $t_s$ is the time when the system collapses. %Then we assume the time dark halo become the Schwarzschild black hole is $t_s$,
%thus the duration of dark halo to black hole is $\Delta t=t_s-t_0$.
%By solving $r^,s$ differential equations, we get the mathematical expression of $\Delta t$ as follows:
%Thus we can calculate the time which the halo mass become the Schwarzschild black hole
From \ref{eq:dEGWdt}, \ref{eq:Eorb},\ref{eq:dEorbdt}, we find
\begin{equation}
\dot{r}=-{64\over 5}{G^3\over c^5r^3}(m_1+m_2)m_1m_2~,
\end{equation}
which yields to
\begin{equation}\label{eq:TBH}
\Delta t={5(r_0^4-r_s^4)c^5\over 256G^3Mm_1m_2}~,
\end{equation}
where $r_0$, $r_s$ are the initial and final radius of the system, respectively.
%while the black hole dome is:
If we take $r_s$ as the Schwarzschild BH radius with mass of the system,
i.e. $r_s={2MG\over c^2}$~,
then the timescale of evolution is completely determined.
Notice that the time for such BH to evaporate is
\begin{equation}\label{eq:THK}
\Delta t_s={5120\pi G^2M^3\over \hbar c^4}~,
\end{equation}
where $ \hbar$ is the reduced Planck constant.

\section{Application to simulation data}\label{Sec:method}
From the above estimation, we see that the timescale depends on the mass and orbital radius change. The cosmic structure formation results in a series of collapsed dark matter halos
whose mass is distributed in a wide range.
These objects would become stable after entering the state of virial equilibrium,
and are expected to stay stable for a long period if no major merger happens.

\begin{figure}
	\centering
	\includegraphics[width=8cm]{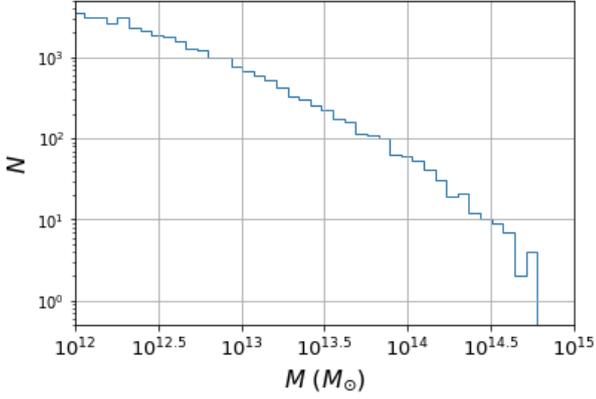}
	\caption{
	Mass function of the halo samples used in this analysis.
        The samples include a set of 56,549 dark matter halos at redshift $z=0$
        in a volume of $(256\ h^{-1} {\rm Mpc})^3$,
        created using N-body method and the \textsc{ROCKSTAR} halo finder.
        }
    \label{fig_massfun}
\end{figure}

\ref{fig_massfun} illustrates the distribution of dark matter halos at redshift $z=0$  with $M>10^{12}M_{\odot}$ in a small volume $(256 h^{-1} {\rm Mpc})^3$,
created by dark matter simulation \citep{GADGET,Tassev13,BD}
with cosmological parameters of
$\Omega_m = 0.307115$, $\Omega_b = 0.048206$, $\sigma_8 = 0.8288$, $n_s = 0.9611$, and $H_0 = 67.77\ {\rm km}\ s^{-1} {\rm Mpc}^{-1}$.
%\cite{Springel:2005mi,Tassev:cola,Koda:2015mca,klypin2016multidark}.

\begin{figure}
	\centering
	\includegraphics[width=8cm]{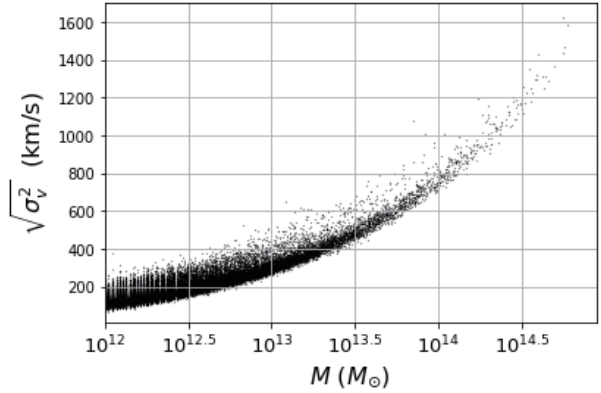}
	\caption{
	Mass-velocity distribution of
       the halo samples used in our analysis.
        %The upper panel shows the results measured in the simulation, while in the lower panel the velocities are calculated
        %using Equations \ref{eq:Virial}\ref{eq:Velocity}.
        }
    \label{fig_Virialtheorem}
\end{figure}

%The data we used is provided by MultiDark Datvenabase,
%which provides results from standard LCDM cosmological simulations.他
%The ``SNAp'' has 56549 lines,
In total we select a sample of 56,549 halos, 
using the \textsc{ROCKSTAR}  halo finder \citep{ROCKSTAR} that allows for robust tracking of substructure based on adaptive hierarchical refinement of friends-of-friends groups
in six phase-space dimensions and one time dimension.
%each line represents a dark matter halo,
%we choose the one whose redshift=0 for our study.
%we get the data of RockStar halo mass of the globular cluster,
%and we choose a cluster which is very consistent with the virial theorem,
%where $\sigma_{r}$ represents velocity dispersion.
In general, the halos we obtain well obey the virial theorem,
which states that a stabilized dynamical system should obey $\frac{GM}{R}\approx\sigma^2_v$,
with $\sigma_v^2$ being the velocity dispersion.
The mass-velocity distribution measured from the simulation is presented in \ref{fig_Virialtheorem}.

To simplify the situation and make everything calculable,
in what follows we will equate the whole dark matter halo to a binary system of two objects with equal mass orbiting around the center, 
and then the virial theorem leads to \ref{eq:Virial}.
%{\bf ? What is $M$ and $m$... Should they not be $M/2$ with $M$ the total mass??}
If we assume the distance between the two objects $R={1\over2} r_0$,
where $r_0$ being the initial radius of the system, then we have
\begin{equation}\label{eq:Virial}
{1\over 2}m(3\sigma^2_{v})\sim{1\over 2}{GMm\over ({1\over 2}r_0)} ~,
\end{equation}
where $3\sigma^2_{v}$ is the $3-D$ velocity squared in the radial direction
and $M,\ m$ both equal to half of the halo mass.
%The upper panel of Figure \ref{fig_Virialtheorem} is the mass-velocity
%distribution of the halo sample used in our work.
Then we have
%\begin{equation}\label{eq:Velocity}
${\sigma^2_{v}={2GM\over 3r_0}}~,$ %to approximate the relation.
%\end{equation}
%To verify the validity of the halo sample, we derive the velocity
%in terms of Equation \ref{eq:Virial} as follows:
%In the lower panel of Figure \ref{fig_Virialtheorem}  we plot
which is a good approximation for what we have measured from the N-body simulation.
%Thus, Equation \ref{fig_Virialtheorem} is a good approximation
%for computing the velocity distribution of a sample based on its mass
%and radius.
%As we can see the distribution of upper and lower panels are almost the same,
%and the distribution of upper panel is more scattered,
%这里的more scattered和下面那句有因果关系吗
%implying that the halo sample is quite reasonable through the comparison.

% \begin{figure}%[htbp]
% 	\centering
% 	%\includegraphics[width=9.4cm]{fig1_a-eps-converted-to.pdf}
% 	\includegraphics[width=8cm]{Mass-Velocity.eps}
% 	%\includegraphics[width=8cm]{virialMass-Velocity.eps}
% 	\caption{
% 	Mass-velocity distribution of
%        the halo samples used in our analysis.
%         %The upper panel shows the results measured in the simulation, while in the lower panel the velocities are calculated
%         %using Equations \ref{eq:Virial}\ref{eq:Velocity}.
%         }
%     \label[figure]{fig_Virialtheorem}
% \end{figure}

As consequences of GW emission, we believe that as time elapses,
the radius of the system $r$ would shrink,
the orbital velocity $v$ would increase, and
the magnitudes of potential energy $V$ and kinetic energy $T$ would both increase.
The gradual changes of the physical quantities are rather small
at the beginning, and dramatically increase at the end of the evolution.

\begin{figure*}%[htbp]
 \centering
 \includegraphics[width=15.5cm]{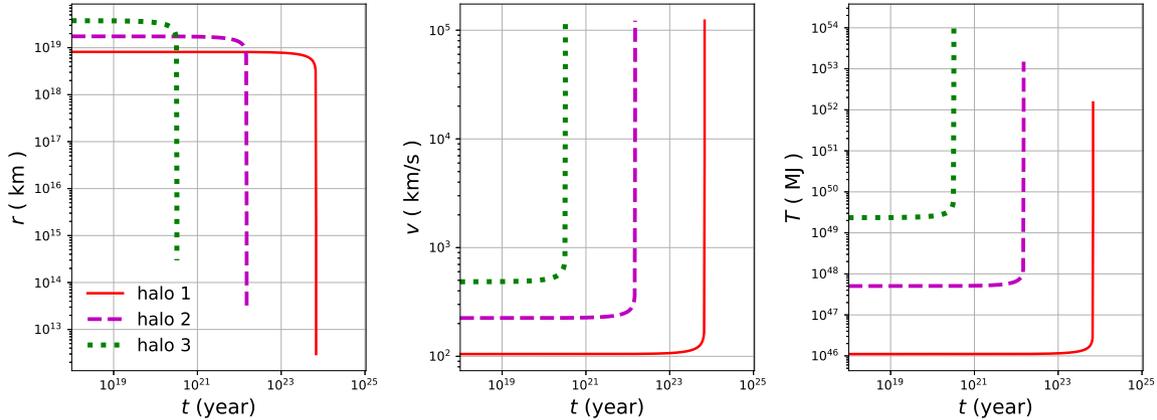}
 \caption{ The evolution of three dynamical systems
 %Their radius, velocities, three
 %radius $r\sim 5.7e+16km, 3.2e+17km, 1.8e+8km$,
 %velocity $v\sim 1040km/s, 1383km/s, 1842km/s$,
 %absolute value of potential energy $V\sim 1.1e+48MJ, 1.9e+49MJ, 3.4e+50MJ$
 %as well as kinetic energy $T\sim 3.0e+24MJ, 5.4e+24MJ, 9.5e+24MJ$.
 %(see \ref{halodata} for detailed data),
 selected from the halo sample (see \ref{tab1} for detailed data). %in terms of the minimum values of
 %$|M_i-10^{12}M_{\odot}|$,$|M_i-10^{13}M_{\odot}|$ and $|M_i-10^{14}M_{\odot}|$ separately,
 %where i is the subscript of the halo sample which is sorted by quality from small to large.
 With time elapsing, their radius $r$ shrinks,
 while the velocity $v$ and kinetic energy $T$ increase.
 The evolution becomes rapid at the end of the evolution.}
 \label{fig:evolution}
\end{figure*}

\ref{fig:evolution} shows the evolution of three typical systems selected from the sample,
having mass around $10^{12}$, $10^{13}$ and $10^{14}$ $M_{\odot}$, respectively.
Their circular velocity and kinetic energy are presented in \ref{tab1}.
%radius $r=5.72852035956612e+16(km), 3.160320829107119e+17(km), 1.7684272434958625e+18(km)$,
%velocity $v=1040.138399686181(km/s)$, $1382.6298072322602(km/s)$, and $1842.2500877147115(km/s)$,
%absolute value of potential energy $V=1.1309246470410093e+48(MJ)$, $1.947966677097805e+49(MJ)$, and $3.4356445755260643e+50(MJ)$
%as well as kinetic energy $T=3.041289047512359e+24(MJ)$,$5.37387139387308e+24(MJ)$, and $9.540532381048152e+24(MJ)$,
% radius $r\sim 5.7e+16km, 3.2e+17km, 1.8e+18km$,
% velocity $v\sim 1040km/s, 1383km/s, 1842km/s$,
% absolute value of potential energy $V\sim 1.1e+48MJ, 1.9e+49MJ, 3.4e+50MJ$
% as well as kinetic energy $T\sim 3.0e+24MJ, 5.4e+24MJ, 9.5e+24MJ$.
% (see \ref{halodata} for detailed data)
% 把数据写成约等于，然后具体数值列成表格会不会美观一点？
% 做了个表格，如果觉得可以的话直接取消掉注释就好啦
 \begin{table*}	
     \begin{center}
         \label{halodata}
         \caption{Dynamical properties of the three selected halos at the present time \label{tab1}
         {\bf }}
         \begin{spacing}{1.3}
         \footnotesize
 		\setlength{\tabcolsep}{3.3mm}{	 	
 		\begin{tabular}{cccccc}
 			\toprule
 			$i$  & Mass $M_{\odot}$ & Radius of the system  $(\rm km)$ &  Circular Velocity $(\rm km/s)$ & Kinetic Energy $(MJ)$  \\
 			\hline
 			1 & $1.01 \times10^{12}$ & $8.15 \times10^{18}$ &  86.4 &$ 7.53 \times10^{45}$ \\  	
 			\hline
 			2 & $1.00\times10^{13}$ & $1.75 \times10^{19}$ &  197.8 &$ 3.90 \times10^{47}$\\
             \hline
			3 & $1.00\times10^{14}$ & $3.77 \times10^{19}$ &  584.8 &$3.41 \times10^{49}$  \\
             \bottomrule
         \end{tabular}}
         \end{spacing}
     \end{center}
 \end{table*}
%which are selected from the halo sample in terms of the minimum values of
%$|M_i-10^{12}M_{\odot}|$,$|M_i-10^{13}M_{\odot}|$ and $|M_i-10^{14}M_{\odot}|$ separately,
%where i is the subscript of the halo sample which is sorted by quality from small to large.
We find that the results are consistent with what we expected.
As demonstrated in \ref{fig:evolution},
with time elapsing, the radius of the system $r$ shrinks,
velocity $v$ and kinetic energy $T$  increase,
and the changes are rapid at the end of the evolution.

% Applying the calculation to the whole sample,
% we get the results of Equation \ref{eq:TBH} of the halos turning into BHs via GW radiation
% and Equation \ref{eq:THK} of the BHs to evaporate through Hawking radiation.

%By applying the whole sample to the calculation of \ref{eq:TBH} and  \ref{eq:THK},
%we obtain the halo distribution in terms of
%the timescale of the halos evolving into BHs via GW radiation, and that of
%the BHs to evaporate through Hawking radiation. %, as shwon in Figure \ref{fig:tste}.
From \ref{eq:TBH} and  \ref{eq:THK}, we obtain the timescale distribution of halo-size binaries to evolve into BHs via GW radiation and BHs evaporation through Hawking radiation.  

\begin{figure*}%[htbp]
\centering
 %\includegra\includegraphics
 %\includegraphics[width=7cm]{t_s_1.eps}
 %\includegraphics[width=7cm]{t_e_1.eps}
\includegraphics[width=14cm]{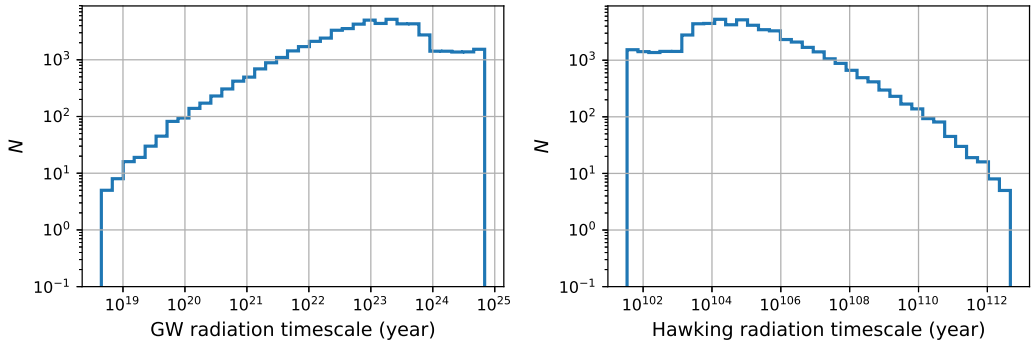}
 \caption{The left panel shows the time needed for the halos to evolve to BHs through GW radiation.
 Time for these BHs to evaporate via Hawking radiation is demonstrated in the right panel.}
 \label{fig:tste}
\end{figure*}

% 或者还要加上什么其他的内容吗？
The left panel of the \ref{fig:tste} illustrates halo distribution
with the time period of GW radiation, which is estimated as $\sim 10^{19}-10^{25}$ years.
%The maximum time we obtain in years is in $10^{26}$ orders.
The right panel shows that the time for BHs to evaporate is much longer,
which distributes in $\sim10^{100}-10^{112}$ years.

\section{Caveats and Discussions}\label{Sec:caveats}

In the above section, the typical timescale that dark matter systems
take to emit GWs before ceasing evolution is estimated using a two-body model.
While the model is oversimplified, our result is still representative and makes physical sense due to two reasons. First, no matter how complicated a many-body system is, as long as the motions (velocities, accelerations) of the objects within are comparable to those in a two-body system with similar mass, size and kinetic energy, the evolution of these two systems should not differ from each other significantly. %So our estimation is roughly correct.
Secondly, the two-body system is the only case of a gravitational system that is everlastingly stable. It is possible that most complicated systems would finally evolve to a stage similar to a two-body system after a long enough period.

However, besides the two-body simplification we made in the calculation, %two-body simplification we make,
there are more caveats that could potentially invalidate our results.
We will summarize and discuss them as follows.

%In what follows we discuss some issues which may affect the picture of the cosmic thermal death besides our simplified model.
%\cite{}

\subsection{Dynamical Evolution, Chaos, Stability}

%## Chaos in Galaxy

%From https://arxiv.org/pdf/astro-ph/9601091.pdf:
%1204.0709
%\cite{Muzzio:2012pe}

%The role that chaos played in this process is not clearly known.
%First of all, the influence of chaos in galaxies are not negligible.
Using the N-body methods, one can start with a certain distribution of
mass points, and trace their evolution by integrating
the equations of motion \citep{AB1978}.
%If one begins with, e.g.,
%a spherical distribution of mass points with very small velocities,
%gravity will force the collapse of
%such system and the radial orbit instability will lead it toward a triaxial equilibrium
%distribution (see, e.g., \citep{Merritt1993}),
%while the obtained triaxiality depends on the initial velocity dispersion.
%## Central black hole effect
%An issue we should pay attention to is the effect of central black hole.
%Merritt and Quinlan (1998) built an N-body model of traxial
%galaxy and verified its stability with a black hole growing in the center.
%The existense of black hole makes the central density
%distribution much steeper, and leads to a loss of triaxiality.
%Different results were demonstrated by Poon \& Merrit (2002, 2004)
%who obtained equilibrium models in which the central regions of traxial
%galaxies contain black holes and are stable.  (persistent for many crossing time).
%## From wiki, N-body problem:
%(http://cncc.bingj.com/cache.aspx?q=+Chaos+in+Gravitational+N-Body+Systems&d=4962323684590480&mkt=en-US&setlang=en-US&w=MJNPlLjV--TQmoVd6lzf7jXDbOw8IyAB)
%The limitations of the N-body method
Yet we are not clear whether this kind of  
stability can sustain within a timescale of $10^{19}-10^{25}$ years due to many well-known difficulties,
%Even though for a small number of bodies, an n-body problem can be
%solved using direct methods, also called particle-particle methods,
%which numerically integrate the differential equations of motion.
%Numerical integration for this problem can be a challenge for several reasons.
e.g., singular gravitational potential when two particles become too close to each other,
and chaotic behavior of the N-body problem when $N>2$.
% limit our ability in studying this problem.
%The gravitational potential is singular.
%It goes to infinity as the distance between two particles goes to zero.
%So the gravitational potential has to be softened to remove the singularity at small distances,
%which may make the results not valid after such long time evolution. %:[18]
%    U ε = ∑ 1 ≤ i < j ≤ n G m i m j ‖ q j − q i ‖ 2 + ε 2 . {\displaystyle U_{\varepsilon }=\sum _{1\leq i<j\leq n}{\frac {Gm_{i}m_{j}}{\sqrt {\left\|\mathbf {q} _{j}-\mathbf {q} _{i}\right\|^{2}+\varepsilon ^{2}}}}.} {\displaystyle U_{\varepsilon }=\sum _{1\leq i<j\leq n}{\frac {Gm_{i}m_{j}}{\sqrt {\left\|\mathbf {q} _{j}-\mathbf {q} _{i}\right\|^{2}+\varepsilon ^{2}}}}.}
%Moreover, in general for N larger than 2 the N-body problem is chaotic, %[36]
%so small errors in the computation may grow exponentially in time.
Although some studies \citep{Voglis2002,
Kalapotharakos2005,Kalapotharakos:2007gn,Muzzio:2012pe}
have shown that it is possible to obtain models
of elliptical galaxies which are %that contain very high fractions of chaotic orbits
highly stable over time intervals of
the order of a Hubble time (even with a black hole growing in the center; see \cite{MQ1998,PM2002,PM2004}),
%The traxial galaxy can be stable
%even with a black hole growing in the center
%Nevertheless,
so far it is impossible to trace
the evolution of the system
within the very long period of the GW radiation.
%Numerical errors will accumulate and increase
%A simulation may be over large stretches of model time
%and numerical errors will accumulate as integration time increases.

\subsection{Dark matter life-time}

%From https://arxiv.org/pdf/1202.1454.pdf:
%

Possibility exists that dark matter consists of
particles that actually decay on a very long time
scale.
%This issue has been studied since a long time,
%e.g. in the context of gravitino dark matter with R-parity violation [191].
\cite{Pandey:2019cre, Mukherjee:2019seu}  showed that dark matter with decay lifetime $\sim 10^{28}-10^{29} $ seconds
can help to explain some detected excesses \citep{Chen:2008dh, Yin:2008bs,Ishiwata:2008cv,Ibarra:2008jk,Chen:2008qs,Nardi:2008ix,Arvanitaki:2008hq},
while some other work \citep{Cohen:2016uyg,Cadena:2019lor} suggested that
dark matter should be stable within a long timescale of $\sim10^{24}-10^{28}$ seconds.
These timescales are already close to the GW emission timescale
($\sim10^{26}-10^{32}$ in unit of second).
%For more studies about the dark matter life time, one can refer to
%[1201.1454 \textbf{can not be found, substituted with 1202.1454 in the following citation}, 1201.5902, 1909.00222, 1910.05017, 1910.11158, 1911.03191]
%\cite{Cirelli:2012tf}
%and the references therein.\underline{}
 %\cite{Cirelli:2012tf}

\subsection{Dark matter accretion by central black hole}

%In galaxy clusters, the existence of the central blak hole
%makes us worry that it may keep accerate the dark matter particles
%and terminate the whole dynamical system. % after a long enough time.
%\footnote{This is supported by the analysis of the bolometric quasar luminosity function at different redshifts,
%which indicates that most of the accreted mass by black holes
%should be barynoic in origin \citep{Soltan:1982vf,1992MNRAS.259..725S,Hopkins_2007}.}
%[8] A. Soltan, MNRAS 200, 115 (1982)\cite{Soltan:1982vf}
%[9] T.A. Small and R.D. Blanford, MNRAS 259, 725 (1992)\cite{1992MNRAS.259..725S}
%[10] P.F. Hopkins, G.T. Richards and L. Hernquist, ApJ 654,731 (2007)\cite{Hopkins_2007}
%On the other hand,
There have been scenarios envisaged in which dark matter could give a significant contribution to the mass accretion of the central black hole of the galaxies.
\cite{PhysRevLett.88.101301} showed that black holes with mass $\ge 10^6 M_{\odot}$ could be formed directly
as a consequence of relativistic core collapse of halos
if halos embedding galaxies are constituted of self-interacting dark matter \citep{Dave_2001},
while \cite{Zelnikov:2003du} and \cite{Munyaneza:2004ai} showed that a significant accretion may happen in the case when dark matter particles are
scattered by stars in molecular clouds near the central black hole.
Although in general the accretion of collisionless dark matter particles into black holes
are thought to be less efficient than that expected from the dissipative baryonic fluid \citep{Peirani:2008bu},
current studies can not rule out the possibility that the dynamical system
would be significantly affected or even terminated due to the black hole accretion
in the long timescale of GW emission.
%can not answer the question that,
%how the dynamical system is going to be affected by the black hole accretion
%in the very long period  of the GW emission stage.
%For more studies on this topic one can refer to e.g.
%\cite{Hu:2005cd, Umeda:2009yc, Guzman:2011ka, Pepe:2011dk, Lora-Clavijo:2014kha, Gupta:2015xok, Eby:2018zlv,East:2019dxt}.
%However,

\subsection{Quadrupole moment in halos}

To be precise, the rate of GW emission depends on the changing rate of the quadrupole moment \citep{Eistein1916,Eistein1918} of the system,
i.e. 
\begin{equation}\label{eq:quadrupole}
{d\over dt} E_{GW}={32\over 5}{G\over c^5}I^2e^2\omega^6 ~,
\end{equation}
where $I$ is the moment of inertia of the halo, 
$e$ is the eccentricity of the halo, and $\omega$ is the angular velocity. 
Using the above formula, we re-calculate the GW emission efficiency of the halo sample
using the positions and velocities of their member particles,
and find the result is actually $34 \pm 15$ times larger than the result obtained 
using the simple two-body model adopted in this analysis.
In any case, these two results do not differ for many orders of magnitude. Therefore our main conclusion remains valid.
%62954 halos composed of particles are used to calculate its quadrupole moment on the rate of gravitational wave emission. 
%The value of quadrupole moment is obtained by the equation \ref{eq:quadrupole}, 
%where the 
%Finally, it is compared with the rate of gravitational wave obtained by our two-body model \citep{PhysRevLett.116.061102}. 
%The result shows that the rate of the gravitational wave radiation of our simulation is about $0.722\%$ to $217.4\%$ of 
%the quadrupole moment calculated by the 62594 halos which is equal to $34.135 \pm 14.60$, 
%here, the mean of the ratio is $34.135$ and the standrd of the ratio is $14.60$. 
%The result is not much different from our result. 
%In consideration of the uncertainty of the future, we still adopt the conclusion of our two bodies as the basis of our research.

The calculation of the quadrupole moment is based on the current mass distribution within the halo samples,
and it is very likely that this mass distribution will change dramatically within the long period of GW emission.
Thus, in this paper we still adopt the simple two-body model to estimate the power of the GW emission.

\subsection{Angular momentum loss, Dynamic friction and the Final parsec problem}

Although halos may lose their angular momentum through interactions with the other objects,
we believe that this will not have significant effect on our main conclusion.
As the universe evolves, the shrinking of the horizon size will isolate halos,
and prohibit their interactions with each other. 
The above process is expected to happen in a timescale much shorter 
than the GW emission timescale.

In particular, the future event horizon, i.e., the region that light emitted
from an object can reach in the infinite future, takes the form of 
\begin{equation}
 r_{\rm FEH}\equiv \int_{t=t_{\rm emit}}^{+\infty} \frac{cdt}{a}.
\end{equation}
If the $\Lambda$CDM is adopted as the theoretical model,
the future event horizon would shrink to $1$ Mpc within $\sim10^{12}$ years, 
which is much shorter compared with the GW emission timescale.

Another process that may contribute to the collapse is the dynamic friction.
In general, its timescale depends inversely on the mass of the system \citep{Chandrasekhar1942}. 
%Once the collapse of a cluster is completed, 
Once a cluster is collapsed, violent relaxation is ineffective, 
and further relaxation occurs through two-body interactions.
\cite{Schechter:1976,Schechter1976} calculated the drag force and the dynamic friction timescale of
a system composed by homogeneous, isotropic, Maxwellian distributed particles.
Following their calculation, we find that the dynamic friction of typical galaxy clusters 
is about $2 \times 10^8$ year, which is also far less than the GW emission timescale.

Finally, one may wonder whether our results is related with the {\it final parsec problem}.
For that, our opinions are as follows.
On one hand, if there exists any mechanism that may fasten the merger of supermassive binary black holes (SBBHs) when they are extremely close to each other, 
such mechanism may also fasten the final stage of the evolution of a halo to a BH. 
In this case, the thermal death may occur a bit earlier.
On the other hand, considering that the timescale of the GW emission is $10^{19} - 10^{25}$ years, 
it is extremely difficult to predict the status of the dynamic system after such a long time evolution. 
One possible situation is that, since two-body system is the only stable gravitational system, 
by the time all halos would exist in a form of two-body system. 
Without a third object interacting with them, there maybe no mechanism that can accelerate the evolution of the system.

%In this paper, the merger of a binary SMBH is not considered, the problem of the supermassive black holes are extremely close is not the case in our current study, but the final parsec problem is a important problem which is neeeded for more attention.

\subsection{Other Fates of the universe}

%The last big unknown is, the nature of the dark energy and
%the role it is going to play in the fate of the universe.
In the case that the nature of dark energy is phantom like ($w<-1$, $\dot \rho  >0$),
its density will reach infinity in a finite
time, disrupt any bounded system, and cause
a "big rip" fate of the universe
\citep{Caldwell:2003vq,Li:2012via},
making our analysis completely meaningless.
Similarly, our analysis is also invalid
in the scenario of bouncing cosmology (see \cite{yifu2014} and the references therein).
%if that is going to happen.

\section{Conclusion}\label{Sec:conclusion}

In this paper, we study the role of GW in the heat death.
GWs happen in almost all dynamical systems in the universe,
causing persistent mechanical energy dissipation in the long-term evolution of the universe and driving the universe to "a state of universal rest and death".
With the N-body simulations, we adopt a simple yet representative scheme
to compute the GW emission process,
and estimate its timescale as $\sim10^{19}-10^{25}$ years,
which depends on the mass of the system. %and then to evaporate.
This timescale is significantly larger than any baryon process in the universe,
but still $\sim10^{80}$ times shorter than the timescale of Hawking radiation.

%As a whole picture, the system emit GW and dissipate all the time,

%it plays an important role in the energy loss of dark matter who will influence the ultimate fate of the universe.

%We use a two-body model to describe globular cluster, i,e. the halo cluster and the emitted energy's effect on the system.
%Aided by simulations we are able to calculate
%and statistically study the the time of halos to BHs via GW radiation and the time of BHs to evaporate via Hawking radiation.
%The timescale of halo to BHs via GW radiation is $\sim10^{25}$yr, and the BHs to evaporate via Hawking radiation is  $\sim10^{111}$yr.
%We find the timescale for the process of halos to BHs is smaller than the process of BHs to evaporate.

By taking the GW emission into consideration,
our work extends the scope of heat death.
This means that, similar to the baryon systems, dark matter has to lose mechanical energy via radiation 
and therefore cannot persist forever.

In this work, we study typical halos with mass $\sim10^{12}-10^{15} M_{\odot}$.
But the results may be different if using a simulation with higher resolution and bigger size,
or if the nature of dark matter is not as assumed in the standard cold dark matter scheme.

%and through the simulation we get the timescale of halos to BHs via GW radiation.
%The significantly  That means dark matter systems can not persist forever as a result of the GW radiation.
%Similar to the baryons Like baryons emit ...,{\bf Don't quite understand}
%they emit,...
%and can not avoid the fate of entering a ``rest state''.

Base on our current knowledge and technology, it is unlikely to obtain a comprehensive understand of 
the physics process of dynamical chaos, dark matter decays or annihilations, central black hole accretion, and the nature of dark energy.
Neither do we know the roles they are going to play in the GW emission process.
Thus, our analysis would be invalid under certain circumstances.

We adopt a rather simplified model to estimate the timescale of the GW emission.
A further study may require running a N-body simulation with the GW emission considered,
which is of great challenge and, not that necessary,
given that there are so many unknown factors that may affect the process.

Several issues about the gravitational wave and thermal death are not discussed in this short work.
In a generic anisotropic universe without positive cosmological constant,
it is possible to generate an infinite amount of entropy
by taking advantage of cosmological shear and curvature anisotropy,
allowing civilizations to exist forever \citep{Far_future,Barrow:2003}.
This provides a novel method to avoid thermal death by utilizing the gravitational waves.
Neither do we discuss the "gravitational field entropy" \citep{Barrow:1986TheAC,Vihan:2017}, 
which is associated with the increase of irregularity in the universe 
and provides a novel picture of the universal heat death.

In all, heat death is an interesting conjecture about the fate of the universe,
and we expect more studies on this issue with
the progress of our understanding about the physics processes in the universe.

%The simulation for the typical timescale that dark matter systems take to emit GWS using a simpler two-body model.
%It is idealized by assuming the systems is in equilibrium, while there exits more caveats which can make our results no longer valid.
%the physics of dynamical chaos, central black hole, dark matter decay and dark energy.
%Include dynamical evolution and chaos, dark matter life-time and dark matter accretion as well as dark energy and fate of the universe.
%Where n-body problem is chaotic, we need a accurate simulation model.
%We should think about that what is the real situation of a dark matter and the influencing factors of the dark matter life,
%as well as the relationship between them.
%Extremely, We need to consider the nature of dark energy and
%the role it plays in the fate of the universe.

%Heat death is a conjecture fate of th universe, and it is still in debate nowdays.
%Our work is just a small part of the study of the heat death, but we enlarge the scope
%by discussing the role of {\it gravitational waves} in it and we get a timescale of the halos to BHs via {\it gravitational waves},
%it has a certain significance. But there exist many caveats need to be discussed.
%We expect more and more discoveries and ideas for this study with progress of physics development.

\section{*Data Availability}

The data used to support the findings of this study are available from the corresponding authors upon request.

\section*{Acknowledgements}

We thank Prof. Rongxin Miao for helpful discussions.
We acknowledge the use of {\it Kunlun} cluster located in School of Physics and Astronomy, Sun Yat-Sen University.
This work is supported by National SKA Program of China No. 2020SKA0110401. 
XDL acknowledges support from the NSFC grant (No. 11803094),
the Science and Technology Program of Guangzhou, China (No. 202002030360).

\bibliographystyle{aasjournal}

\end{document}